\documentclass[prd,superscriptaddress,nofootinbib,showpacs,preprint]{revtex4}
\usepackage{graphicx}
\usepackage[usenames,dvipsnames]{color}
\usepackage{amsmath,amssymb}
\usepackage[colorlinks]{hyperref}
\newcommand{\vev}[1]{\langle #1 \rangle}
\begin{document}
\title{Spontaneous Left-Right Symmetry Breaking in Supersymmetric Models
with only Higgs Doublets}
\author{Sudhanwa Patra}
\email{sudhakar@prl.res.in}
\affiliation{Physical Research Laboratory, Ahmedabad-380009, India}
\author{Anjishnu Sarkar}
\email{anjishnu@iopb.res.in}
\affiliation{Institute Of Physics, Bhubaneswar-751005, India}
\author{Utpal Sarkar}
\email{utpal@prl.res.in}
\affiliation{Physical Research Laboratory, Ahmedabad-380009, India}

\begin{abstract}

We studied the question of parity breaking in a supersymmetric
left-right model, in which the left-right symmetry is broken with
Higgs doublets (carrying $B-L=\pm 1$). Unlike the left-right
symmetric models with triplet Higgs scalars (carrying $B-L=\pm
2$), in this model it is possible to break parity spontaneously by
adding a parity odd singlet. We then discussed how neutrino mass
of type III seesaw can be invoked in this model by adding extra
fermion singlets. We considered simple forms of the mass matrices
that are consistent with the unification scheme and demonstrate
how they can reproduce the required neutrino mixing matrix. In
this model, the baryon asymmetry of the universe is generated via
leptogenesis. The required mass scales in the model is then found
to be consistent with the gauge coupling unification.

\end{abstract}
\maketitle

\section{Introduction}
\label{sec:intro}
The existence of massive neutrinos, the unknown origin of parity
violation in the Standard Model (SM) and the hierarchy problem are
some of the important motivations for physics beyond the SM.
The most natural extension of the standard model that addresses
these issues is the supersymmetric version of the left-right
symmetric extension of the standard model, which will treat
the left-handed and right-handed particles on equal footing,
and the parity violation we observe at low energies
would be due to the spontaneous breaking of the left-right symmetry at
some high scale \cite{Pati:1974yy,Mohapatra:1974gc,Sahu:2006pf,
Senjanovic:1975rk,Mohapatra:1980qe,Babu:1998wi,Babu:1992ia,Babu:2008ep,Deshpande:1990ip}.
Another interesting feature of the
left-right symmetric model is that the difference between the
baryon number (B) and the lepton number (L) becomes a gauge symmetry,
which leads to several interesting consequences.

In spite of the several virtues of the minimal supersymmetric
left-right symmetric models (MSLRM), we are yet to arrive at a
fully consistent model, from which we can descend down to the MSSM.
One of the most important problems is the spontaneous breaking of
left-right symmetry \cite{Kuchimanchi:1993jg,Kuchimanchi:1995vk}.
There has been suggestions to solve this problem by introducing
additional fields or higher dimensional operators or by going
through a different symmetry breaking chain or breaking the
left-right symmetry around the supersymmetry breaking scale
\cite{Kuchimanchi:1993jg,Kuchimanchi:1995vk,Babu:2008ep,Aulakh:1998nn,
Sarkar:2007er, Aulakh:1997vc, Aulakh:1997fq}. In some cases,
this problem is cured through the introduction
of a parity-odd singlet, but the soft susy breaking terms
then lead to breaking of
electromagnetic charge invariance. One interesting SUSYLR
model is the minimal SUSYLR model,
which has been studied extensively \cite{Aulakh:1997ba,
Kuchimanchi:1993jg, Kuchimanchi:1995vk},
and it has been found that
global minimum of the Higgs potential is either charge violating or
$R$-parity violating.

Recently we proposed
yet another solution to the problem, which resembles
the non-supersymmetric solution, relating the vacuum expectation
values (vev's) of the left-handed and right-handed triplet
Higgs scalars to the Higgs bi-doublet vev through a seesaw
relation. We achieved this by introducing a bi-triplet and singlet
Higgs scalars, and the vacuum that preserves both
electric charge and R-parity can naturally be the global minimum of the
full potential. In this article we are applying this idea of
spontaneous left-right symmetry breaking at high scale in
supersymmetric models with only doublet Higgs scalars. We
extend the model with one singlet Higgs scalar,
which breaks the left-right parity of the gauge groups
at a high scale.
The most attractive feature of the present model
is that it does not allow any left-right symmetric solution to be
a minimum of the potential. We also discuss the question
of neutrino masses via type III see-saw mechanism and
leptogenesis in details. We then embed the model in a grand
unified theory and study the gauge coupling unification to
check the consistency of the mass scales required in this
model.

\section{Minimal SUSYLR model: a brief review}
In this section, we shall review the left-right extension of the
standard model, where the gauge group at higher energies
is the left-right symmetric group $G_{LR}\equiv SU(3)_C \times SU(2)_L \times SU(2)_{R} \times
U(1)_{B-L}$ and we assume that at energies above the TeV scale,
the theory is supersymmetric. In these supersymmetric left-right
symmetric models, it is assumed that
the MSSM gauge group $SU(3)_C \otimes SU(2)_L \otimes
U(1)_Y$ is enhanced at some higher energy, above which the left-handed
and right-handed fermions are treated on equal footing.
The minimal supersymmetric left-right (SUSYLR) model starts with
the left-right symmetric gauge group $G_{LR}$,
which could emerge from a supersymmetric $SO(10)$ grand unified theory.
The field content of this model is given by,
\begin{eqnarray}
Q = (3,2,1,1/3), &\quad& Q^c = (3,1,2,-1/3), \nonumber \\
L = (1,2,1,-1), &\quad& L^c = (1,1,2,1),
\end{eqnarray}
where the numbers in the brackets denote the quantum numbers under
$G_{LR}$.

The Higgs sector of this model consists of the bidoublet and triplet
superfields, given by,
\begin{eqnarray}
\Phi_i = (1,2,2,0), &\quad& (i = 1,2), \nonumber \\
\Delta = (1,3,1,2), &\quad& \bar{\Delta} = (1,3,1,-2), \nonumber \\
\Delta^c = (1,1,3,-2), &\quad& \bar{\Delta}^c = (1,1,3,2).
\end{eqnarray}
Under the left-right parity corresponding to the interchange of the
gauge groups $SU(2)_L$ and $SU(2)_R$, or the
D-parity, the fields transform as
\begin{eqnarray}
Q \leftrightarrow {Q^*}^c, &\quad& L\leftrightarrow {L^*}^c, \nonumber \\
\Delta \leftrightarrow {\Delta^*}^c, &\quad&
\bar{\Delta} \leftrightarrow \bar{\Delta}^{\ast\,c}, \nonumber \\
\Phi_i\leftrightarrow
\Phi_i^{\dagger}.
\end{eqnarray}
The superpotential for this theory is given by
\begin{eqnarray}
W &=& \nonumber Y^{(i)_{q}} Q^{T} \tau_{2} \Phi_{i} \tau_{2} Q^{c}
  + Y^{(i)_{l}} L^{T} \tau_{2} \Phi_{i} \tau_{2} L^{c}\\
\nonumber
&& + ~i (f L^{T} \tau_{2} \Delta L +f^{*} L{^{c}}^{T} \tau_{2}
  \Delta^{c} L^{c} ) \\
&& + ~\mu_{\Delta} \textrm{Tr}(\Delta \bar{\Delta}) + \mu^{*}_{\Delta}
  \textrm{Tr}(\Delta^{c} \bar{\Delta}^{c})+ \mu_{ij}
  \textrm{Tr}(\tau_{2}\Phi^{T}_{i} \tau_{2} \Phi_{j}).
\end{eqnarray}
One of the important problems with the supersymmetric left-right
extension of the standard model is that the minimization of the
potential does not allow spontaneous parity breaking, which was
considered to be one of the major triumph of the
non-supersymmetric LR models. Several attempts were made to solve
this problem in some variants of the model.
Some of these solutions involve
modifying the Higgs sector, adding higher dimensional operators or
involving a different breaking scheme of the group theory
\cite{Aulakh:1997ba,Aulakh:1997fq,Aulakh:1998nn, Sarkar:2007er,Aulakh:1997vc,Patra:2009wc}.
The simplest solution is to include a bi-triplet field \cite{Patra:2009wc} and
allow D-parity breaking at some high scale, which may then
allow parity violation spontaneously, allowing the scale of
$SU(2)_R$ breaking to be different from the $SU(2)_L$ breaking
scale. We extend that argument to the models involving only
doublets.

In models with only doublet scalars, we require three singlet
fermions to give masses to the neutrinos. The charged fermion
masses originate from the vev of the bi-doublet scalar field.
Since there are no triplet scalar field that breaks the
symmetry $SU(2)_R$, and all the triplet scalars are replaced
by the doublet scalars, the bi-doublet field required to give
masses to the charged fermions can give rise to the coupling
required to break the parity, when D-parity is broken. Thus
left-right symmetry breaking becomes more natural in these
models with only doublet scalar fields. \color{black}
This model is also able to generate baryon asymmetry via
leptogenesis and provide neutrino masses through both
inverse see-saw mechanism and also
using Type-III see-saw mechanism.
\section{SUSYLR with Higgs doublets and parity odd singlet}
We consider here a SUSYLR model with only doublet Higgs scalars, which
is the simplest extension of the non-supersymmetric LR model. This includes
the bi-doublet scalar field that is required to give masses to the charged
fermions and also to break the $SU(2)_L$ symmetry after the left-right
symmetry is broken.
The doubling of the bidoublet
Higgs in previous models was to ensure a non-vanishing CKM matrix. For the
sake of simplicity of our model we forgo this condition since it doesn't
have any bearing on parity breaking. However, extension of the present model
via doubling of the bidoublet is fairly trivial. Thus, the Higgs sector of
our model is given by,
\begin{eqnarray}
\chi_L \equiv (1,2,1,-1),
&\quad&
\bar{\chi}_L \equiv (1,2,1,1),
\nonumber \\
\chi_R \equiv (1,1,2,-1),
&\quad&
\bar{\chi}_R \equiv (1,1,2,1),
\nonumber \\
\Phi = (1,2,2,0), &\quad& \sigma \equiv (1,1,1,0).
\end{eqnarray}
where, with usual custom the subscript $L$ and $R$ denotes the left and
right handedness of the Higgs particle. The Higgs particles with ``bar" in
the notation, helps in anomaly cancellation of the model.

We have also
included a singlet scalar field $\sigma$, which has the special property
that it is even under the usual parity of the Lorentz group,
but it is odd under the
parity that relates the gauge groups $SU(2)_L$ and $SU(2)_R$.
This field $\sigma$ is thus a scalar and not a pseudo-scalar field,
but under the D-parity transformation that interchanges $SU(2)_L$
with $SU(2)_R$, it is odd. This kind of work is proposed in
\cite{Chang:1984qr,Hirsch:2006ft}. Although all the scalar fields are
even under the parity of the Lorentz group, under the D-parity the
Higgs sector transforms as,
\begin{eqnarray}
\chi_L \leftrightarrow \chi_R, &\quad&
\bar{\chi}_L \leftrightarrow \bar{\chi}_R,
\nonumber \\
\Phi \leftrightarrow \Phi^\dagger, &\quad& \sigma \leftrightarrow -\sigma.
\end{eqnarray}
The superpotential of the model relevant in the context of parity breaking
is given by,
\begin{eqnarray}
W & = & f \Phi \left( \bar{\chi}_{L} \chi_{R}
+ \chi_{L}\bar{\chi}_{R} \right) + m_{\Phi} \Phi \Phi
\nonumber \\ &&
~ + m_{\chi} \left( \bar{\chi}_{L} \chi_{L}
+ \bar{\chi}_{R} \chi_{R} \right)
\nonumber \\ &&
~ + m_\sigma \sigma^2 + \lambda \sigma (\bar{\chi}_L \chi_L
- \bar{\chi}_R \chi_R).
\end{eqnarray}
Supersymmetry being unbroken, implies the $F$ and $D$ conditions are equal
to zero. The $F$ flatness conditions for the various Higgs fields are
given by,
\begin{eqnarray}
F_{\Phi} &=& f \left( \bar{\chi}_{L} \chi_{R}+\chi_{L}\bar{\chi}_{R} \right)
+2 \,m_{\Phi} \Phi = 0,
\nonumber \\
F_{\chi_{L}} &=& f \Phi \bar{\chi}_{R} +m_{\chi} \bar{\chi}_{L}
+ \lambda \sigma \bar{\chi}_L = 0,
\nonumber \\
F_{\bar{\chi}_{L}} &=& f  \Phi \chi_{R}+m_{\chi} \chi_{L}
+ \lambda \sigma \chi_L = 0,
\nonumber \\
F_{\chi_{R}} &=& f \Phi \bar{\chi}_{L} +m_{\chi} \bar{\chi}_{R}
- \lambda \sigma \bar{\chi}_R = 0,
\nonumber \\
F_{\bar{\chi}_{R}} &=& f  \Phi \chi_{L}+m_{\chi} \chi_{R}
- \lambda \sigma \chi_L = 0,
\nonumber \\
F_\sigma &=& 2 m_\sigma \sigma + \lambda (\bar{\chi}_L \chi_L
- \bar{\chi}_R \chi_R).
\end{eqnarray}
Similarly, the $D$ flatness conditions, are given by,
\begin{eqnarray}
D_{R_i} &=& \chi_R^\dagger \tau_i \chi_R
+ \bar{\chi}_R^\dagger \tau_i \bar{\chi}_R = 0, \nonumber \\
D_{L_i} &=& \chi_L^\dagger \tau_i \chi_L
+ \bar{\chi}_L^\dagger \tau_i \bar{\chi}_L = 0, \nonumber \\
D_{B-L} &=& (\chi_L^\dagger \chi_L - \bar{\chi}_L^\dagger \bar{\chi}_L )
- (\chi_R^\dagger \chi_R - \bar{\chi}_R^\dagger \bar{\chi}_R ) = 0.
\end{eqnarray}
In both the $F$ and $D$ flat conditions we have neglected the lepton
fields, since they would have a zero vev. The vev's for the scalar fields
are given by,
\begin{eqnarray}
\vev{\chi_{L}} &=& \vev{\bar{\chi}_{L}} = v_{L},
\nonumber \\
\vev{\chi_{R}} &=& \vev{\bar{\chi}_{R}} = v_{R},
\nonumber \\
\vev{\Phi} &=& v, \quad \vev{\sigma} = s.
\end{eqnarray}
Here, for simplicity of the model, we have assumed $\chi_L$ and
$\bar{\chi}_L$ to have the same vev $v_L$. Similarly, for the right-handed
fields $\chi_R$ and $\bar{\chi}_R$.

Minimization of $D$ flat conditions, leads to a number of holomorphic gauge
invariants which corresponds to flat directions \cite{Luty:1995sd}.
Here, however, in order to determine the vacuum structure of our model, we
minimize the $F$ flat conditions and discuss about the relations that emerge
from them.

After the scalar fields have acquired their respective vevs, the $F$
flatness conditions are given by,
\begin{eqnarray}
F_\Phi &=& f(v_{L} v_{R}+v_{R} v_{L})+2 m _{\Phi} v = 0,
\label{eq:Fphivev} \\
F_{\chi_{L}} &=& f v v_{R}+\lambda s v_{L} +m_{\chi} v_{L} = 0,
\label{eq:FchiLvev} \\
F_{\bar{\chi}_{L}} &=& f v v_{R}+\lambda s v_{L} +m_{\chi} v_{L} = 0,
\label{eq:FchiLbvev}\\
F_{\chi_{R}} &=& f v v_{L}-\lambda s v_{R} +m_{\chi} v_{R} = 0,
\label{eq:FchiRvev}\\
F_{\bar{\chi}_{R}} &=& f v v_{L} -\lambda s v_{R} +m_{\chi} v_{R} = 0,
\label{eq:FchiRbvev}\\
F_\sigma &=& 2 m_{\sigma}s +\lambda(v_{L}^{2}-v_{R}^{2}) = 0.
\label{eq:Fsigmavev}
\end{eqnarray}
Solving the equations we get four relations among the vevs.
\begin{eqnarray}
v_{L} &=& \frac{-m_{\Phi} v}{f v_{R}}
\label{eq:phi_vev}  \\
m_{\chi}+ \lambda s &=& \frac{f\, v \,v_R}{v_L}
\label{eq:chiL_vev} \\
m_{\chi}-\lambda s &=& -\frac{f v v_{L}}{v_{R}}
\label{eq:chiR_vev} \\
s &=& \frac{\lambda}{2 m_{\sigma}}(v_{R}^{2}-v_{L}^{2})
\label{eq:sigma_vev}
\end{eqnarray}
The role of D-parity odd singlets $\sigma$ is uni-important in left-right
breaking. This can be understood from eqns. (\ref{eq:chiL_vev})
and (\ref{eq:chiR_vev}) as follows:
\begin{equation}
\left( \frac{v_L}{v_R} \right)^2 =
\frac{M-\lambda\, s}{M+\lambda \, s}
\label{eq:ratio}
\end{equation}
If there is no $\sigma$ field, then $s=0$. This implies $v_L=v_R$ which is a
left-right symmetric solution. Also the F-term conditions
(\ref{eq:FchiLvev})-(\ref{eq:FchiRbvev}) are not consistent without the
inclusion of the parity odd singlet $\sigma$ in the model. Hence, the
parity odd singlet $\sigma$ is necessary to account for the spontaneous
left-right breaking and for the consistency of the model.

We now try to interpret these results to get a working phenomenology.
Considering the last of the relations eqn (\ref{eq:sigma_vev}) we see that
$s = 0$ is a trivial solution, and will put $v_L$ and $v_R$ on equal
footing thus leading to unbroken parity. However, $s = 0$ is a
special solution of eqn (\ref{eq:sigma_vev}). For $s \neq 0$, we have
$v_L \neq v_R$ and parity is violated spontaneously. We will choose
$v_R \gg v_L$, as it is usually assumed in model building for
phenomenological reasons. Choosing the mass ($m_\Phi$) and vev ($v$) of
$\Phi$ to be of electroweak (EW) scale and considering the dimensionless
coupling constant $\lambda$ to be of order unity, we immediately come to the
conclusion, from eqn (\ref{eq:chiR_vev}), that $m_\chi \sim s$.

In order to avoid generic susy problems like over abundance of gravitino,
we assume the mass scale of $v_R$ to be $\le 10^9$ GeV. This together with
eqn (\ref{eq:phi_vev}) gives the value of $v_L \simeq 10^{-5}$ GeV, where
$f$, another dimensionless quantity, without any fine-tuning is considered
to be of order unity. This is also consistent with the assumption that
$v_R \gg v_L$. Now using eqn (\ref{eq:chiL_vev}) and the above derived
relation that $m_\chi \sim s$ we get $m_\chi \sim s \simeq 10^{16}$ GeV.
Finally, from eqn (\ref{eq:sigma_vev}) one derives the mass of $\sigma$
($m_\sigma$) to be of EW scale.
\begin{table}[htb]
\begin{center}
\begin{tabular}{|c|c|}
\hline
Masses/Vevs & Case - I (In GeV) \\
\hline
$m_\chi, s$ & $10^{16}$\\
$v_R$ & $10^9$ \\
$m_\Phi, v$ ,$m_\sigma$ & $10^2$\\
$v_L$ & $10^{-5}$ \\
\hline
\end{tabular}
\caption{Mass scales of the model}
\label{tab:mass_scale}
\end{center}
\end{table}
If one considers non-thermal leptogenesis, then one can consider the
alternative possibility of having a low value of $v_R$ i.e.
$\sim \mathcal{O}(10)$ TeV. Then all the mass scales and vevs are reduced by
a couple of orders and could be accessible to colliders. The results are
summarized in Table (\ref{tab:mass_scale}).
\section{Neutrino mass and leptogenesis in SUSYLR model with Higgs doublet}
In LR models with only doublet scalar fields, the question
of neutrino masses and leptogenesis has been discussed in details. We shall
try to restrict ourselves as close as possible to these existing non-supersymmetric
models, and check the consistency of these solutions when parity is broken
in the present SUSYLR model. We shall first discuss the scenario with
conserved D-parity, but since LR symmetry cannot be broken without breaking
D-parity we shall discuss the D-parity breaking scenario afterwards.

In conventional type I seesaw, neutrino mass can be realized via three right
handed neutrinos $N^{c}_{i}$ where we have Majorana mass term
$(M_{R})_{ij} N^{c}_{i}N^{c}_{j}$ and Dirac masses with the ordinary
neutrinos $(M_{N})_{ij} \nu_{i} N^{c}_{j}= (Y_{N})_{ij}\nu_{i} N^{c}_{j}
\langle \Phi \rangle$. After diagonalizing, the resulting neutrino mass is
$M^{I}_{\nu}=-M_{N}\, M_{R}^{-1}\, M_{N}^{T}$. Type II seesaw requires a
$SU(2)_{L}$ triplet Higgs field $T$ with mass of order $m_{T}$.
Integrating out the Higgs triplet $T$ leads to an mass operator
$(M_{T})_{ij} \nu_{i}\nu_{j}$ with $M_{T} \propto \frac{Y_{T}\langle \Phi
\rangle^{2}}{m_{T}} \sim \frac{v^{2}}{M_{G}}$. Combination of these neutrino
mass are also possible in left-right models which contains both type I and
type-II or, type I and type III \cite{Ma:1998dn,Bajc:2006ia}.

In type III neutrino mass \cite{Foot:1988aq} three hypercharge neutral
fermionic triplets $\Sigma^a ~(a=1,2,3)$ are added to explain the $\nu$
mass term. In our model, however, we have an extra fermionic superfield
which give rise $\nu$ mass term which is similar to the conventional
type III seesaw mechanism. Thus, it is in this spirit that we can call
the seesaw mechanism in our model as type III seesaw. For the review of the
standard type III seesaw mechanism we closely follow
\cite{Albright:2003xb}.

Along with the Dirac neutrino mass term $(M_N)_{ij} \nu_i N^c_j$, the
relevant superpotential for $\nu$ mass term, which is due to the extra
fermion singlet $(S)$ is given by,
\begin{equation}
W = M_{ij} S_{i} S_{j}+F_{ij} {l_{L}}_{i} S_{j} \chi_{L}
+ F'_{ij} {l_{R}}_{i} S_{j} \chi_{R},
\label{eq:spot}
\end{equation}
From the above superpotential one can see that the vev of the left-handed
doublet Higgs field which acquires a low scale vev $\vev{\chi{L}} = v_{L}$
directly couples the left-handed $\nu_i's$ with the singlet $S_i$.
The mass matrix for the neutral leptons has the form,
\begin{equation}
W_\textrm{neut} = (\nu_i \quad  N^c_i \quad  S_i)
        \left( \begin{array}{ccc}
                0        & (M_N)_{ij}   & F_{ij} v_{L} \\
              (M_N)_{ji} &    0         & F_{ij} v_{R} \\
              F_{ji} v_{L} & F_{ji} v_{R} & {\cal M}_{ij}
        \end{array} \right)
        \left( \begin{array}{c}
            \nu_j \\ N^c_j \\ S_j
        \end{array} \right).
\label{eq:spotmatrix}
\end{equation}
In the above mass matrix, the mass of the singlet $\mathcal{M}_{ij}$ and
the vev of the right-handed Higgs doublet $v_R$ are heavy, while $M_N$ and
vev of the left-handed Higgs doublet $v_L$ are of low scale.

Since in our model we have more than one left-handed Higgs doublet
$(\chi_L, \bar{\chi}_R)$, the $\nu$ mass is given by,
\begin{eqnarray}
M_{\nu} &=& - M_N M_R^{-1} M_N^T - (M_N H + H^T M_N^T)
    \left(\frac{v_{L}}{v_{R}}\right),
\label{eq:typeIII}
\\
\textrm{where,}\qquad
H &\equiv& \left( F' \cdot F^{-1}\right)^{T},
\label{eq:Hmatrix}
\\
M_{R} &=& (F \,v_{R}) {\cal M}^{-1} (F^T v_{R}).
\label{eq:calM}
\end{eqnarray}
The first term in eqn (\ref{eq:typeIII}) is the type I seesaw contribution
and the second term gives the type III seesaw contribution.
Type III contribution to $\nu$ mass will dominate over type I if the
elements of the matrix $\mathcal{M}_{ij}$ are small compared to the
contribution of $H$ term.

We will partly follow the formalism and parametrization used in
\cite{Albright:2003xb,Albright:2004ws} where the elements of the Dirac
mass matrix are ${M_{N}}_{11}=\eta v$, ${M_{N}}_{33}=v$, ${M_{N}}_{23} =
-{M_{N}}_{32} = v \epsilon$ and else are zero. Here $\eta=0.6\times10^{-5}$
and $\epsilon \sim 0.14 $.

If the elements of $F_{ij}$ and $F'_{ij}$ are considered to be of the order
of $f$, a dimensionless parameter then from eqn. (\ref{eq:Hmatrix}) we find
that $H_{ij} \sim 1$ $(i,j=1,2,3)$. Thus, the $\nu$ mass resulting from
eqn (\ref{eq:typeIII}) is
\begin{equation}
 M_{\nu} =
\left(\begin{array}{ccc}
 \eta    & \epsilon & 1 \\
\epsilon & \epsilon & 1 \\
   1     & 1        & 1
\end{array}\right)\frac{v\, v_{L}}{v_{R}}
\label{eq:nmass}
\end{equation}
The neutrino mass as presented above mostly satisfy the observed neutrino
mass with a minor fine tuning in the $13$ element.

Another set of parameters can be chosen to explain both neutrino mass and
leptogenesis where both $F_{ij}$ and $F'_{ij}$ take the form
\cite{Albright:2003xb}
\begin{equation}
F,F' \sim \left( \begin{array}{ccc} \lambda^2 & \lambda & \lambda \\
\lambda & 1 & 1 \\ \lambda & 1 & 1 \end{array} \right),
\label{eq:ff}
\end{equation}
where $\lambda \sim \eta/\epsilon$.
With this form of $F,F'$ we have from eqns (\ref{eq:typeIII}) and
(\ref{eq:ff}),
\begin{equation}
H \sim \left( \begin{array}{ccc}
1 & \epsilon/\eta & \epsilon/\eta \\ \eta/\epsilon & 1 & 1 \\
\eta/\epsilon & 1 & 1 \end{array} \right),
\end{equation}
and
\begin{equation}
M_{\nu} \sim \left( \begin{array}{ccc} \eta & \epsilon & \epsilon \\
\epsilon & \epsilon & 1 \\ \epsilon & 1 & 1 \end{array}
\right) \frac{v \,v_L}{v_R}.
\end{equation}
For the study of leptogenesis, a diagonal $F_{ij}$ would suffice better.
The parameters in this new basis would be represented via a tilde.
The right-handed neutrino and the singlet has to be transformed via a
unitary transformation to attain the diagonal basis as such $N^c_i = U_{ij}
\tilde{N}^c_j$ and $S_i = V_{ij} \tilde{S}_j$. To attain the diagonal form
of $F_{ij}$ the unitary matrix $U_{ij}$ can have the form
\begin{equation}
U = \left( \begin{array}{ccc}
    u_{11} & \lambda u_{12} & \lambda u_{13} \\
    \lambda u_{21} & u_{22} & u_{23} \\
    \lambda u_{31} & u_{32} & u_{33}
    \end{array} \right)
\end{equation}
with $V_{ij}$ having a similar form. Here the $u_{ij}$ elements are of
$\mathcal{O}(1)$. For simplicity and numerical computation we will use the
particular form of the unitary matrix which is
\begin{equation}
U = \left(
\begin{array}{ccc}
1 &  -\lambda (1 +\sqrt{2})i &  \lambda \\
-\lambda (1+\sqrt{2})i & 1/\sqrt{2} & i/\sqrt{2} \\
 \lambda  & i/\sqrt{2} &1/\sqrt{2}
\end{array} \right).
\end{equation}
The elements of the diagonalized matrix $\tilde{F}_{ij} v_R = (U_{ki}
F_{k \ell} V_{\ell j}) v_R$ can be written
\begin{equation}
\tilde{F} v_R = \textrm{diag} [\lambda^2 F_1, F_2, F_3] v_R
    \equiv \textrm{diag}[M_1, M_2, M_3],
\end{equation}
where $F_i \sim 1$. In this basis the matrices $\tilde{F}'_{ij} u$ and
$\tilde{{\cal M}}_{ij}$ can be parametrized as
\begin{eqnarray}
\tilde{F}' u &=& \left( \begin{array}{ccc}
    \lambda^2 f_{11} & \lambda f_{12} & \lambda f_{13} \\
    \lambda f_{21} & f_{22} & f_{23} \\
    \lambda f_{31} & f_{32} & f_{33}
    \end{array} \right) v,
\nonumber \\
\tilde{{\cal M}} &=& \left( \begin{array}{ccc}
    \lambda^2 g_{11} & \lambda g_{12} & \lambda g_{13} \\
    \lambda g_{21} & g_{22} & g_{23} \\
    \lambda g_{31} & g_{32} & g_{33}
    \end{array} \right) M_S,
\label{eq:Mtilde}
\end{eqnarray}
where, $f_{ij}, g_{ij} \sim 1$. The assumption here is that the scale of
$M_S \ll v_R$.
In the new basis, the Dirac neutrino mass matrix $M_N$ transforms
as $\tilde{M}_N = M_N U$ and the form of the transformed matrix is
\begin{equation}
\label{eq:yukawa}
\tilde{M}_N \cong \left( \begin{array}{ccc}
\eta u_{11} & \eta \lambda u_{12} & \eta \lambda u_{13} \\
\epsilon \lambda u_{31} & \epsilon u_{32} & \epsilon u_{33} \\
\lambda u_{31} & u_{32} & u_{33} \end{array} \right) v \equiv
\tilde{Y} v.
\end{equation}
After doing all the parametrization, the type III seesaw contribution to
the light neutrino mass matrix (which dominates, since $M_S \ll v_R$) from
eqn (\ref{eq:typeIII}) is given by,
\begin{equation}
M_{\nu} \cong - \left[ \begin{array}{ccc}
2 \eta \left( \frac{u_{11} f_{11}}{F_1} \right) & \frac{\eta}{\lambda}
\left( \frac{u_{11} f_{21}}{F_1} \right) & \frac{\eta}{\lambda}
\left( \frac{u_{11} f_{31}}{F_1} \right) \\ \frac{\eta}{\lambda}
\left( \frac{u_{11} f_{21}}{F_1} \right) & 2 \epsilon \sum_j
\left( \frac{u_{3j} f_{2j}}{F_j} \right) &
\sum_j \left( \frac{u_{3j} f_{2j}}{F_j} \right) \\
\frac{\eta}{\lambda} \left( \frac{u_{11} f_{31}}{F_1} \right)
& \sum_j \left( \frac{u_{3j} f_{2j}}{F_j} \right) &
2 \sum_j \left( \frac{u_{3j} f_{3j}}{F_j} \right) \end{array}
\right] \left( \frac{v^2}{v_R} \right).
\end{equation}

Now we discuss the leptogenesis scenario in the given form of the neutrino
matrix $M_N$, ${\cal M}$, $M_S$ and $U$ \cite{Albright:2003xb,Albright:2004ws}. Consider the case where the six
super heavy two-component neutrinos have the mass matrix
\begin{equation}
(\tilde{N}^c_i, \tilde{S}_i) \left( \begin{array}{cc}
0 & M_i \delta_{ij} \\ M_i \delta_{ij} & \tilde{{\cal M}}_{ij} \end{array}
\right) \left( \begin{array}{c} \tilde{N}^c_j \\ \tilde{S}_j \end{array}
\right),
\end{equation}
where, $\tilde{{\cal M}}_{ij}$ is given in eqn (\ref{eq:Mtilde}).
The leptogenesis can be realized by the decays of the lightest
pair of these super heavy neutrinos, which have effectively the $2\times2$
mass matrix
\begin{equation}
(\tilde{N}^c_1, \tilde{S}_1) \left( \begin{array}{cc}
0 & M_1  \\ M_1  & \tilde{{\cal M}}_{11} \end{array}
\right) \left( \begin{array}{c} \tilde{N}^c_1 \\ \tilde{S}_1 \end{array}
\right) =
(\tilde{N}^c_1, \tilde{S}_1) \; \lambda^2 \left( \begin{array}{cc}
0 & F_1 v_R  \\ F_1 v_R  & g_{11} M_S \end{array}
\right) \left( \begin{array}{c} \tilde{N}^c_1 \\ \tilde{S}_1 \end{array}
\right).
\end{equation}
Consider the scenario where $M_S \ll v_R$, then this results an almost degenerate
pseudo-Dirac pair or equivalently two Majorana neutrinos with nearly equal
and opposite masses. These Majorana neutrinos are
$N_{\pm} \cong (\tilde{N}^c_1 \pm \tilde{S}_1)/\sqrt{2}$, with masses
$M_{\pm} \cong \pm M_1 + \frac{1}{2} \tilde{{\cal M}}_{11} =
\lambda^2 (\pm F_1 v_R + \frac{1}{2} g_{11} M_S)$.
These can decay into light neutrino plus Higgs via the term
$Y_{i \pm}(N_{\pm} \nu_i)H$, where
\begin{equation}
\label{eq:yukawalepto}
Y_{i \pm} \cong (\tilde{Y}_{i1} \pm \tilde{F}'_{i1})/\sqrt{2}
\mp \frac{\tilde{{\cal M}}_{11}}{4 M_1}
(\tilde{Y}_{i1} \mp \tilde{F}'_{i1})/\sqrt{2}.
\end{equation}
Here $\tilde{Y}$ is the Dirac Yukawa coupling matrix given in eqn (\ref{eq:yukawa}).
It is straightforward to show that the lepton asymmetry produced by the decays of
$N_{\pm}$ \cite{Albright:2003xb} is given by
\begin{equation}
\label{eq:asynot}
\epsilon_1 = \frac{1}{4 \pi} \frac{Im [\sum_j(Y_{j+} Y^*_{j-})]^2}
{\sum_j [|Y_{j+}|^2 + |Y_{j-}|^2]} I(M^2_-/M^2_+),
\end{equation}
where $f(M^2_{1+}/M^2_{1-})$ comes from the absorptive part of the decay
amplitude of $N_{\pm}$ . This function is given by
\begin{equation}
\label{eq:fact}
I(x)=\sqrt{x}\left[\frac{1}{1-x}+1-(1+x)\, {\rm ln}\left(\frac{1+x}{x}
\right) \right]
\end{equation}
Making use of eqns (\ref{eq:yukawalepto}) and (\ref{eq:asynot}) one obtains
\begin{eqnarray}
\epsilon_1 &=& \frac{1}{4 \pi} \frac{\sum_j(|\tilde{Y}_{j1}|^2 -
|\tilde{F}'_{j1}|^2) {\rm Im}(\sum_k \tilde{Y}^*_{k1} \tilde{F}'_{k1})}
{\sum_j (|\tilde{Y}_{j1}|^2 + |\tilde{F}'_{j1}|^2)} f(M^2_{1+}/M^2_{1-}),
\nonumber \\
\textrm{or, } \qquad \epsilon_1 &\cong& \frac{\lambda^2}{4 \pi}
    \left[ \frac{(|u_{31}|^2 - |f'_{31}|^2) {\rm Im}(u_{31}^* f'_{31})}
    {|u_{31}|^2 + |f'_{31}|^2 + |f'_{21}|^2}
    \right]
    f(M^2_{1+}/M^2_{1-}).
\label{eq:asy}
\end{eqnarray}
\begin{table}[t]
\begin{tabular}{|l|c|c|c|c|}
\hline
Input & Case (III-1)\qquad & Case (III-2)\qquad & Case (III-3)
         \qquad & Case (III-4)\qquad\\[0.1in] \hline\hline
$v_R$ (GeV) & $2.7 \times 10^{14}$ & $2.7 \times 10^{12}$ & $8.8\times 10^{10}$ & $9.8 \times 10^{8}$ \\[0.07in]
$F_1$ & 1.0 & 10. & 31 & 50 \\[0.07in]
$F_2$ & 1.0 & 0.1 & 0.1 & 1.0 \\[0.07in]
$F_3$ & 1.0 & 1.0 & 1.0 & 1.0 \\[0.07in]
$M_S$(GeV) & $4.3 \times 10^5$ & $430$ & $43$ & 10.0 \\[0.07in]
$f_{21}$ & -0.950 + 0.534i & -0.050 + 0.0534 i & -0.950 + 0.11 i & -0.01+0.01 i \\[0.07in]
$f_{22}$ & -2.279 - 1.537i & -0.227 - 0.154i & -0.228 - 0.154i & -0.225+0.138 i \\[0.07in]
$f_{23}$ & -0.194 + 1.523i & -0.194 + 1.523i & -0.193 + 0.573 i & -0.195 + 1.23 i \\[0.07in]
$f_{31}$ & 0.6+3.5 i & -0.012 + 0.385 i  &  -0.46 + 0.42 i   & 0.04 +0.04 i\\[0.07in]
$f_{32}$ & -0.354i   & -0.035i  & -0.035i &  0.023 i   \\[0.07in]
$f_{33}$ & 0.354  & 0.354  & 0.354  & 0.523 \\[0.07in]
\hline
\end{tabular}
\caption{Type III seesaw and Leptogenesis results for four cases}
\label{tab:input}
\end{table}
The lepton asymmetry produced by the decay on lightest Majorana neutrino is
partially diluted by the lepton number violating decay processes. This decay
processes try to wash out the lepton asymmetry already produce before. This
wash out factor is given by,
\begin{equation}
\label{eq:washout}
k(\tilde{m}_1) \sim 0.3 \left( \frac{10^-3 \,\text{eV}}{\tilde{m}_1} \right)
 \left( \log\frac{\tilde{m}_1}{10^-3 \,\text{eV}} \right)^{-0.6}
\end{equation}
The equilibrium mass of the neutrino is given by
\begin{equation}
\label{eq:eqmass}
\tilde{m}_1 \equiv \frac{8 \pi v_u^2 \Gamma_{N_{1\pm}}}{M^2_{N1_{\pm}}}
\cong \lambda^2 \frac{v_u^2}{M_1} (|u_{31}|^2 + |f'_{31}|^2 + |f'_{21}|^2).
\end{equation}
\begin{table}[t]
\begin{tabular}{|l|c|c|c|c|}
\hline
Output & Case (III-1)\qquad & Case (III-2) \qquad & Case (III-3)
         \qquad & Case (III-4) \qquad \\[0.1in] \hline\hline
 $M_1$ (GeV) & $ 4.53 \times 10^5 $ & $ 4.53 \times 10^3$
    & $ 4.58 \times 10^3$ & 82.37\\[0.07in]
 $M_2$ (GeV) & $ 2.70 \times 10^{14}$ & $ 2.70 \times 10^{12}$
    & $ 8.8 \times 10^{10}$ & $9.8 \times 10^8$ \\[0.07in]
 $M_3$ (GeV) & $ 2.70 \times 10^{14}$ & $ 2.70 \times 10^{12}$
    & $8.8 \times 10^{10}$ &$9.8 \times 10^8$ \\[0.07in]
 $(M_{1+} + M_{1-})/M_{1+}$ & $1.6 \times 10^{-9}$ & $1.59 \times 10^{-10}$
    & $1.57 \times 10^{-10}$ & $4.08 \times 10^{-9}$ \\[0.07in]
 $\epsilon_1$ & $-2.5 \times 10^{-6}$
    & $-2.1 \times 10^{-4}$ & $-1.01 \times 10^{-6}$ & $-1.01 \times 10^{-4}$\\[0.07in]
 $\tilde{m}_1$ (eV) & 0.511 & 0.569 & 4.774 &   0.694 \\[0.07in]
 $\kappa_1$ & $5.1 \times 10^{-4}$ & $4.5 \times 10^{-4}$
    & $4.5 \times 10^{-5}$ & $3.6 \times 10^{-4}$\\[0.07in]
 $\eta_B$ & $1.11 \times 10^{-10}$
    & $1.147 \times 10^{-10}$ & $3.911 \times 10^{-10}$ & $1.461 \times 10^{-10}$ \\
\hline
\end{tabular}
\caption{Type III seesaw results for four cases}
\label{tab:output}
\end{table}
\subsection{Numerical Result}
The lepton asymmetry produced per unit entropy, taking into account decays of Majorana
neutrino and their washout factors, is given by
\begin{eqnarray}
\label{eq:leptasy11}
\frac{n_L}{\it{s}} &\cong& \nonumber \frac{k \, \epsilon_1}{\it{s}} \frac{g_N \, T^3}{\pi^2}\\
&\cong &\frac{45}{2\, \pi^4} \frac{g_N}{g_*} k\,\epsilon_1
\end{eqnarray}
We have used the expression for entropy of the comoving volume,
$\it{s}=\frac{2}{45} \,g_* \pi^{2} T^3$. Here $g_N=2$ for Majorana spin degrees freedom and
$g_*=228.75$ is the relativistically spin degrees of freedom for supersymmetry.

The corresponding B-L asymmetry per unit entropy is just the negative of $n_L/\it{s}$, since baryon number is conserved in the right-handed Majorana neutrino decays.While $B-L$
is conserved by the electroweak interaction following those decays, the sphaleron processes violate $B+L$ conservation and convert the $B-L$ asymmetry into a baryon asymmetry.The baryon asymmetry for supersymmetric case is
\begin{equation}
 \frac{n_B}{\it{s}}=-\frac{28}{79} \frac{n_L}{\it{s}}
\label{eq:leptsph}
\end{equation}
With the entropy density $\it{s}=7.04\, n_\gamma$ in terms of the photon
density, the baryon asymmetry($\eta_B$) of the Universe, defined by the
ratio $n_B$ of the net baryon number to the photon number, is given in
terms of the lepton asymmetry($\epsilon_1$) and washout parameter (k) by
\begin{equation}
\eta_B=\frac{n_B}{n_\gamma}\cong -0.039 \, k \, \epsilon_1.
\label{eq:baryasy}
\end{equation}
Successful Leptogensis will require that the final result for $\eta_B$ should be order
of $10^{10}$.
where $\lambda = \eta/\epsilon = 4.1 \times 10^{-5}$ as before.

The input parameter given in the table (\ref{tab:input}) which will
determine the small neutrino mass, leptogenesis parameter as output given
in the table (\ref{tab:output}) of our model.

\section{Gauge coupling unification}

   Grand unified theories (GUTs) offer the possibility of unifying the
three gauge groups viz., $SU(3)$, $SU(2)$ and $U(1)$ of the standard
model into one large group at a high energy scale $M_U$. This scale is
determined as the intersection point of the $SU(3)$, $SU(2)$ and $U(1)$
couplings. The particle content of the theory completely determines the
variation of the couplings with energy. Given the particle content of
the theory one can evolve the couplings, determined at low energies, to
determine whether there is unification or not.

In this section we will discuss how one can obtain $SU(3)_C \times SU(2)_L \times SU(2)_R
\times U(1)_{B-L} (g_L = g_R )(\cong G_{2213} )$ intermediate gauge symmetry in R-parity
conserving supersymmetric grand unified theory through one-loop unification of gauge couplings.
Suppose we want to evolve coupling parameter between the scales $M_1$ and $M_2$ (i.e,
$M_1 \le \mu \le M_2$) corresponding to the two scales of physics, then the RGE's depend on
the gauge symmetry and particle content at $\mu = M_1$.
In table (\ref{tab:susylr}), we give the particle content of the model.

\begin{table}[htb]
\begin{tabular}{|c|c|}
\hline
Fields &$SU(3)_c \times SU(2)_L \times SU(2)_R \times U(1)_{B-L}$ \\
\hline
$Q$         & $(3,2,1,+1/3)$      \\
$Q^c$       & $(3^\ast,1,2,-1/3)$ \\
$L$         & $(1,2,1,-1)$        \\
$L^c$       & $(1,1,2,+1)$        \\
$\chi_{L}$  & $(1,2,1,+1)$        \\
$\chi_{R}$  & $(1,1,2,-1)$        \\
$\bar{\chi}_{L}$  & $(1,2,1,-1)$  \\
$\bar{\chi}_{R}$  & $(1,1,2,+1)$  \\
$\Phi_a$   & $(1,2,2,0)$          \\
$S$        & $(1,1,1,0)$          \\
\hline
\end{tabular}
\caption{Field content of the SUSY LR model}
\label{tab:susylr}
\end{table}
For this purpose, we consider the two step breaking of the group $G$ to the
minimal supersymmetric standard model (MSSM) through $G_{3221}$ intermediate
gauge symmetry in the so called minimal grand unified theory.
\begin{center}
\begin{tabular}{ccll}
$G$ & $\stackrel{M_{U}}{\rightarrow}$ & $SU(3)_{c}\times SU(2)_{L}\times
SU(2)_{R}\times U(1)_{(B-L)}$ & $[G_{3221}]$ \\
& $\stackrel{M_{R}}{\rightarrow}$ & $SU(3)_{c}\times SU(2)_{L}\times
    U(1)_{Y}$ & $[G_{321}]$ \\
 & $\stackrel{M_{W}}{\rightarrow}$ & $SU(3)_{c}\times U(1)_{Q}$
 & $[G_{em}]$.
\end{tabular}
\end{center}

\subsection{RGE for SUSYLR model with doublet Higgs}
The couplings evolve according to their respective beta functions.
The renormalization group equations(RGEs) for this model cane be written as
\begin{equation}
\frac{d \alpha_{i}}{d t}=\alpha^{2}_{i} [b_{i}+\alpha_{j} b_{ij}+ O(\alpha^{2}) ]
\end{equation}
where, $t=2\pi\, ln(\mu)$. The indices $i,j=1,2,3$ refer to the gauge
group $U(1)$, $SU(2)$ and $SU(3)$ respectively.

Unlike the D-parity breaking case where the intermediate left-right gauge group has four
different coupling constants as discussed in \cite{Dev:2009aw}, in the present case
$G_{3221}$ has only three gauge couplings, $g_{2L} = g_{2R}$ , $g_{3C}$ , and $g_{BL}$
for $μ \geq M_{R}$. We now write down the RG evolution equation of gauge couplings
upto one loop order which are given below
\begin{eqnarray}
\frac{1}{\alpha_Y(M_Z)}&=&\frac{1}{\alpha_G}+\frac{a_Y}{ 2\pi}\ln \frac{M_R}{ M_Z}
+\frac{1}{10\pi}\left(3a'_{2L} + 2a'_{BL}\right)\ln\frac{M_U}{ M_R}
 ,\nonumber\\
\frac{1}{\alpha_{2L}(M_Z)}&=&\frac{1}{\alpha_G}+\frac{a_{2L}}{ 2\pi}\ln \frac{M_R}{ M_Z}
+ \frac{a'_{2L}}{ 2\pi}\ln\frac{M_U}{ M_R},                                          \nonumber\\
\frac{1}{\alpha_{3C}(M_Z)}&=&\frac{1}{\alpha_G}+\frac{a_{3C}}{ 2\pi}\ln\frac{M_R}{ M_Z}
+\frac{a'_{3C}}{ 2\pi}\ln \frac{M_U}{ M_R}.
\label{rges}
\end{eqnarray}
where $\alpha_G=g_G^2/4 \pi$ is the GUT fine-structure constant and the beta function
 coefficients $a_i$ and $a'_i$ are determined by the particle  spectrum
 in the ranges  from $M_Z$ to $M_R$, and from  $M_R$ to $M_U$,
 respectively.

Here we are using PDG values, $\alpha(M_Z) = 127.9,~\sin^2\theta_W(M_Z)=
0.2312$ , and $\alpha_{3C}(M_Z) = 0.1187$ \cite{Yao:2006px}.
Consider the case where  $SU(2)_R \times U(1)_{B-L}$ breaks down to $U(1)_Y$.
In that case
\begin{equation}
\frac{Y}{2}=I_{3,R}+\frac{B-L}{2}
\end{equation}
The normalized generators are $I_Y=(\frac{3}{5})^{1/2} \frac{Y}{2}$ and
$I_{B-L} = (\frac{3}{2})^{1/2} \frac{B-L}{2}$. Using these, one can write
\begin{equation}
I_Y=\sqrt{\frac{3}{5}} I_{3,R}+\sqrt{\frac{2}{5}}I_{B-L}
\end{equation}
Which implies that the matching of the coupling constant at the scale where
the left-right symmetry begins to manifest itself is given by
\begin{equation}
 \alpha_Y^{-1}=\frac{3}{5} \alpha_{2R}^{-1}+\frac{2}{5} \alpha_{B-L}^{-1}
\end{equation}

\subsection{Result}

\begin{enumerate}
\item
At scale $\mu=M_Z-M_R $,
\begin{equation}
a_Y = 33/5, a_{2L} = 1, a_{3C} = -3,
\end{equation}
\item
At scale $\mu =  M_R - M_U$,
\begin{eqnarray}
& &b^{\prime}_{BL}=16,b^{\prime}_{2L}=b^{\prime}_{2R}=4 , \, \nonumber \\
& &a'_{3C} = a_{3C} =-3.
\end{eqnarray}
\end{enumerate}
\begin{figure}[htb]
 \centering
 \includegraphics[bb=0 0 240 153]{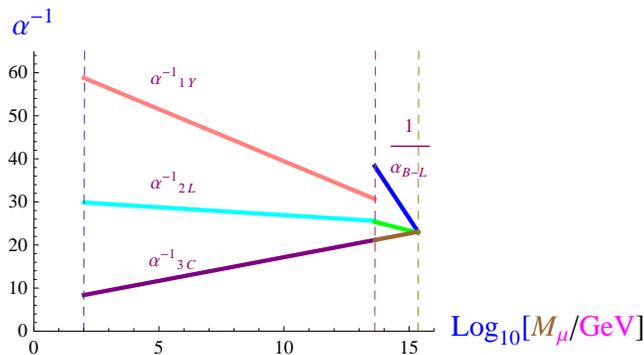}
 \caption{Evolution of coupling constants in susylr model with Higgs doublet}
 \label{fig:susylr}
\end{figure}

This will change once we add contributions coming from extra
particle added to the minimal supersymmetric model. Once we fix the values of
beta functions, we can achieve lower values of $M_{R}$. There are discussion
\cite{Dutta:1998bn,Hempfling:1995rb,Setzer:2005hg}, where the Unification is possible
at the same energy scale around $10^{16}$  GeV, but the scale of
$M_{R}$ varies from $10^{9}$ - $10^{12}$ GeV.

Let us summarize our results. We point out that the non-supersymmetric
version of the Standard Model is ruled out by LEP data.
However, the supersymmetric extension of this scenario remains a viable
alternative to conventional grand unified theories and is capable of
predicting the precision values of couplings determined from LEP and
unification is possible within the error bar. There are model
\cite{Malinsky:2005bi,Dev:2009aw} where one can achieve unification
of all three fundamental interactions in which D-parity is broken at the
GUT level. We see from figure (\ref{fig:susylr}) that the gauge couplings unify at a scale
$5.27\times 10^{15}$ GeV. Also the right handed scale $M_R$ is found to be $2.69\times
10^{13}$ GeV in our model.

\section{Conclusion}

We studied the question of spontaneous parity breaking in the supersymmetric
version of the left-right symmetric models, in which all symmetry breaking
takes place with only doublet Higgs scalars. We demonstrate that unlike the
models with triplet Higgs scalars, in these models the left-right symmetry
could be broken at a different scale compared to the electroweak symmetry
breaking scale, if we introduce a singlet Higgs scalar $\sigma$, which breaks D-parity,
that is the parity relating the gauge groups $SU(2)_L$
and $SU(2)_R$ but not relating to the parity of the Lorentz group. The vev of the
field $\sigma$ breaks the D-parity, but does not break the Lorentz parity.
But when combined with the vevs of the other doublet scalars, it allows to
break the group $SU(2)_R$ at a different scale than the $SU(2)_L$ breaking
scale, which is in the range of  $10^8-10^{13}$ GeV,
(though we can have low $v_R$ which is allowed from minimization of the potential).
We then demonstrated the consistency of the model in terms of
the neutrino mass and the matter-antimatter asymmetry of the
Universe. We then consider embedding of the model and check the consistency
of the mass scales involved for the gauge coupling unification.
\section{Acknowledgments}
Sudhanwa Patra would like to thank Santosh Kumar Singh for useful discussion.
AS would like to thank the hospitality of PRL where most of this work was
developed.


\end{document}